\documentclass[12pt,epsf]{article}
\usepackage{verbatim}

\usepackage{dsfont}
\usepackage{sidecap}
\sidecaptionvpos{figure}{t}
\usepackage{subfigure}
\usepackage{enumerate} 

\usepackage{amssymb,amsmath}
\usepackage{graphicx, xcolor, varwidth}
\usepackage{setspace}
\usepackage[permil]{overpic}
\usepackage{cite}

\colorlet{darkblue}{blue!70!black}
\colorlet{darkgreen}{green!70!black}

\usepackage[colorlinks=true,urlcolor=darkblue,linktocpage=true,linkcolor=darkblue,citecolor=darkblue]{hyperref}

\numberwithin{equation}{section}

\newcommand{\be}{\begin{equation}}
\newcommand{\ee}{\end{equation}}
\newcommand{\bea}{\begin{eqnarray}}
\newcommand{\eea}{\end{eqnarray}}
\newcommand{\bear}{\begin{eqnarray}}
\newcommand{\eear}{\end{eqnarray}}  
\newcommand{\beas}{\begin{eqnarray*}}

\newcommand{\eeas}{\end{eqnarray*}}
\newcommand{\ba}{\begin{array}}
\newcommand{\ea}{\end{array}}

\newcommand{\pd}[2][1]{\ifnum#1=1 \frac{\partial}{\partial {#2}} \else
  \frac{\partial^#1}{\partial {#2}^{#1}}\fi}
\newcommand{\dpd}[2][1]{\ifnum#1=1 \dfrac{\partial}{\partial {#2}} \else
  \frac{\partial^#1}{\partial {#2}^{#1}}\fi}
\newcommand{\td}[2][1]{\ifnum#1=1 \frac{d}{d{#2}} \else
  \frac{d^#1}{d{#2}^{#1}}\fi}

\newcommand{\nbox}{{\,\lower0.9pt\vbox{\hrule \hbox{\vrule height 0.2 cm \hskip 0.19 cm \vrule height 0.2 cm}\hrule}\,}}

\newcommand{\bz}{\bar{z}}



\newdimen\tableauside\tableauside=1.0ex
\newdimen\tableaurule\tableaurule=0.4pt
\newdimen\tableaustep
\def\phantomhrule#1{\hbox{\vbox to0pt{\hrule height\tableaurule width#1\vss}}}
\def\phantomvrule#1{\vbox{\hbox to0pt{\vrule width\tableaurule height#1\hss}}}
\def\sqr{\vbox{%
  \phantomhrule\tableaustep
  \hbox{\phantomvrule\tableaustep\kern\tableaustep\phantomvrule\tableaustep}%
  \hbox{\vbox{\phantomhrule\tableauside}\kern-\tableaurule}}}
\def\squares#1{\hbox{\count0=#1\noindent\loop\sqr
  \advance\count0 by-1 \ifnum\count0>0\repeat}}
\def\tableau#1{\vcenter{\offinterlineskip
  \tableaustep=\tableauside\advance\tableaustep by-\tableaurule
  \kern\normallineskip\hbox
    {\kern\normallineskip\vbox
      {\gettableau#1 0 }%
     \kern\normallineskip\kern\tableaurule}%
  \kern\normallineskip\kern\tableaurule}}
\def\gettableau#1 {\ifnum#1=0\let\next=\null\else
  \squares{#1}\let\next=\gettableau\fi\next}

\newcommand{\smallstart}{
\end{spacing}
\noindent\hfil\rule{1\textwidth}{.4pt}\hfil\small

   \addtolength{\leftskip}{5mm}
}
\newcommand{\smallend}{
   \addtolength{\leftskip}{-5mm}
\noindent\hfil\rule{1\textwidth}{.4pt}\hfil\normalsize
\begin{spacing}{1.3}
}

\renewcommand{\O}{{\cal O}}

\begin{document}
\begin{spacing}{1.3}
\begin{titlepage}
\begin{center}
{\Large \bf 
Conformal Bootstrap Deformations}

\vspace*{6mm}

Nima Afkhami-Jeddi$^1$ 

\vspace*{6mm}

$^1$\textit{Enrico Fermi Institute \& Kadanoff Center for Theoretical Physics,\\ University of Chicago, Chicago, Illinois, USA}

\vspace{6mm}

{nimaaj@uchicago.edu}

\vspace*{6mm}
\end{center}

\begin{abstract}
We explore the space of extremal functionals in the conformal bootstrap. By recasting the bootstrap problem as a set of non-linear equations parameterized by the CFT data, we find an efficient algorithm for converging to the extremal solution corresponding to the boundary of allowed regions in the parameter space of CFTs. Furthermore, by deforming these solutions, we demonstrate that certain solutions corresponding to known theories are continuously connected. Employing these methods, we will explore the space of non-unitary CFTs in the context of modular as well as correlation function bootstrap. In two dimensions, we show that the extremal solution corresponding to the Ising model is connected to that of the Yang-Lee minimal model. By deforming this solution to three dimensions, we provide evidence that the CFT data obtained in this way is compatible with the $\epsilon$-expansion for a non-unitary theory.
\end{abstract}

\setcounter{tocdepth}{1}

\end{titlepage}
\end{spacing}

\vskip 1cm

\setcounter{tocdepth}{1}
\addtocounter{page}{1}

\begin{spacing}{1.3}\section{Introduction}
The conformal bootstrap encompasses a set of analytic and numerical techniques to solve or constrain strongly interacting quantum field theories. These methods have successfully been employed to analytically classify certain theories in two spacetime dimensions \cite{Belavin:1984vu,Polyakov:1974gs}. More recently, numerical methods have been successfully employed in constraining the space of consistent conformal field theories in higher dimensions \cite{Rattazzi:2008pe,Poland:2018epd,ElShowk:2012ht,El-Showk:2014dwa,Poland:2010wg,Kos:2016ysd,Kos:2013tga}. In general, these numerical methods begin with a set of assumptions about the scaling dimensions of operators in the CFT.  One then proceeds by imposing unitarity and crossing equations in order to prove that the set of assumptions leads to a contradiction. By ruling out the starting assumptions we can map out the space of CFTs consistent with unitarity and associativity of the operator product expansion. This task is usually carried out using semidefinite programming\cite{Poland:2011ey}. One then typically tries to find the most stringent set of assumptions\textemdash for example by maximizing the scaling dimension of the first primary scalar above the vacuum\textemdash to find the boundary of the allowed regions where CFTs may exist. It is an interesting observation that certain well-known CFTs are found at or near the boundaries of these allowed regions. Furthermore, at the boundary of the allowed region, one can construct an extremal functional from which additional information about the putative CFT spectrum may be extracted\cite{El-Showk:2012vjm,El-Showk:2016mxr}. These methods have been successfully employed in carving out the parameter space of CFTs in various spacetime dimensions. They have also been used to find islands in parameter space corresponding to the 3d critical Ising model as well as others \cite{Kos:2014bka,Kos:2015mba,Kos:2016ysd,Agmon:2019imm,Chester:2019ifh,Chester:2020iyt}.

Despite their many successes, these numerical methods have certain limitations. One such limitation is the fact that these techniques only work for unitary theories\footnote{See \cite{Gliozzi:2013ysa,Gliozzi:2014jsa,Li:2017ukc,El-Showk:2016mxr,ElShowk:2012hu} for alternative approaches to bootstrap that do not rely on unitarity.}. This is due to the fact that positivity constraints arising from unitarity of the underlying theory play a crucial role in setting up the semideifinite optimization problem. Furthermore, the output of these methods is limited to CFT data corresponding to the low lying operators in the theory. In addition, these methods are computationally expensive and are typically carried out on large computing clusters. 

It was argued in \cite{El-Showk:2016mxr,ElShowk:2012hu} that extremal functional methods provide an alternative approach to the bootstrap problem. They showed that at the boundary of allowed regions in parameter space of CFTs the CFT data satisfies a set equations called extremality conditions. They also showed that by linearizing these equations near a solution, that such solutions may be deformed allowing for flows along the extremal solutions.

Follwing \cite{El-Showk:2016mxr,ElShowk:2012hu}, we will recast problem finding the boundary of the allowed region  as a constrained optimization problem parameterized by the CFT data. This approach has the advantage that it is computationally cheaper. This is achieved by making an ansatz for the extremal functional along the boundary of the allowed region. The choice of ansatz is motivated by the observation that at the boundary, the extremal functional has a constrained form. This observation was leveraged in \cite{Afkhami-Jeddi:2019zci} to construct an algorithm that applies to the spinless modular bootstrap problem introduced by \cite{Hellerman:2009bu}. Although the output of the algorithm is the same as that which is found by employing semidefinite programming, the lowered computational cost of the algorithm resulted in increasing the number of operators one can access by orders of magnitude. The same algorithm was also used to place bounds on optimal sphere packings in large dimensions \cite{Afkhami-Jeddi:2020hde}. The algorithm developed in \cite{Afkhami-Jeddi:2019zci} has the limitation that it applies only to simple problems such as the spinless modular bootstrap as well the one dimensional conformal correlator bootstrap \cite{El-Showk:2016mxr}.

In this paper, we will develop an algorithm that will remedy this problem\footnote{A Mathematica notebook implementing this algorithm for the correlator bootstrap is included with this submission.}. This algorithm can be applied to the more general bootstrap problems in higher dimensions. This involves solving directly for the extremal functional in terms of the CFT data. Although we will not be able to leverage the reduced computational cost of the algorithm to increase the reach of the conformal bootstrap in the same way by increasing the dimension of the functional basis, we will extend the reach of the conformal bootstrap by exploring regions of parameter space not accessible using traditional techniques. This is due to the fact that the algorithm does not require positivity as an input and is therefore able to explore regions corresponding to non-unitary theories. However, this means that although the algorithm produces functionals in the non-unitary regions, they cannot be simply interpreted as defining boundaries of regions where non-unitary theories are allowed to exist\footnote{In some cases, these extremal functionals can still be interpreted as ruling out regions in parameter space of CFTs provided we make additional assumptions on the sign of squared OPE coefficients.}. In addition, since the algorithm is concerned directly with the CFT data, it allows for more control over the specifics of the parameter space we wish to explore.

The rest of the paper is organized as follows: In section \ref{bsasco} we will briefly review the conformal bootstrap and describe how it may be recast as a set of non-linear equations. We will also describe how the methods outlined in this section may be applied to the correlation function as well as the modular bootstrap problems. In section \ref{dts} we describe how solutions to these equations may be deformed in order to explore parameter space of CFTs in 2 and 3 dimensions. We finally conclude with a brief discussion in section \ref{dao}.
\section{Bootstrap as Constrained Optimization} \label{bsasco}
\subsection{Bootstrap review}
In this section, we will briefly review the general bootstrap problem and establish some notations. See \cite{Qualls:2015qjb,Rychkov:2016iqz,Simmons-Duffin:2016gjk,Chester:2019wfx} for more complete reviews of the conformal bootstrap. The starting point for setting up a bootstrap problem is typically a physical quantity that has a convergent expansion in two or more channels.  For example, these channels can correspond to different quantization surfaces in the torus partition function of a 2d CFT resulting in the modular bootstrap or to different choices operator product expansion of a CFT 4-point function on a sphere following from the associativity of the operator algebra. In general, these expansions take the following form\footnote{More generally, this equation depends on the spacetime dimension as well as the quantum number of the external operators appearing in the observables under consideration. We have suppressed such dependence in this expression.}
\begin{align}
\sum_{a \in Q_1}c_a \mathbb{F}_a(\{x_i\})=\sum_{b \in Q_2}\tilde{c}_b \tilde{\mathbb{F}}_b(\{x_i\}),
\end{align}
where $Q_{1,2}$ label the quantum numbers of the exchange operators in each channel, $\mathbb{F}$ and $\tilde{\mathbb{F}}$ are kinematic functions in each of the corresponding channels and whose form is fixed by the spacetime symmetries of the problem under consideration, $c_{a,b}$ are c-numbers corresponding to the degeneracies or the OPE coefficients depending on the nature of the observable under consideration and $\{x_i\}$ represent the coordinates in which the physical quantities are computed. Note that in unitary theories, $c_{a,b} $ are positive numbers. This expression contains infinitely many equations parameterized by $\{x_i\}$ for infinitely many unknowns parameterized by the $c_{a,b} $ and the corresponding quantum numbers in each channel. For simplicity, we will take the observable under question such that $Q_1=Q_2\equiv Q$. This is the case for example when considering the 4-point function of identical operators on a sphere. We can then rewrite this equation as
\begin{align}
\sum_{a \in Q}c_a\mathbb{W}_a(\{x_i\})=0,
\end{align}
where we have defined $\mathbb{W}_a(\{x_i\})\equiv \mathbb{F}_a(\{x_i\})-\tilde{\mathbb{F}}_a(\{x_i\})$. In order to extract useful information from this equation, we will act on this expression by linear functionals $\mathcal{F}^i$. Assuming the action of the functional commutes with the infinite sum we have
\begin{align}\label{funcsum}
\sum_{a \in Q}c_a \mathcal{F}^i [\mathbb{W}_a(\{x_i\})]=0.
\end{align}
It is convenient to choose the functionals to be derivatives with respect to the coordinates $\{x_i\}$ evaluated at symmetric points. However, the choice of functionals is not unique and different choices have been explored with varying degrees of success\cite{Echeverri:2016ztu,Paulos:2019gtx,Mazac:2019shk,Caron-Huot:2020adz,Paulos:2020zxx,Mazac:2018mdx,Mazac:2018qmi,Mazac:2016qev,Mazac:2018ycv,Paulos:2019fkw}.  We will specify the choice of functional for the specific examples that we consider in this paper. Note that the functionals $\mathcal{F}^i_a\equiv \mathcal{F}^i [\mathbb{W}_a(\{x_i\})]$ are functions only of the quantum numbers. Making the assumptions that the quantum number $a$, appearing in the sum \eqref{funcsum} belong to a set $A$, a contradiction is obtained if we manage to set of numbers $\alpha_i$ such that $\alpha_i \mathcal{F}^i_a >0$ for all $a\in A$. In this way, we carve out regions in the parameter space of CFTs by ruling out $A$. We can approach the boundary of the allowed region for example by maximizing over the quantum of numbers of a particular operator appearing in the theory\footnote{See \cite{Paulos:2021jxx} for other choices of optimization that result in approaching boundaries in the parameter space of CFTs.}. Adopting the terminology appearing in \cite{El-Showk:2012vjm} we will refer to $\alpha_i \mathcal{F}^i_a$ as the extremal functional if it corresponds to the boundary of allowed $A$. This maximization is typically carried out by semidefinite programming. Here, we will instead proceed by solving directly for the extremal functional in terms of the CFT data. We will give specific examples of how this can be applied to a variety of bootstrap problems in the rest of this section.
\subsection{Warm-up: spinless modular bootstrap}
To elucidate the logic of this approach, we will consider the spinless modular bootstrap, which is arguably the simplest non-trivial example of a bootstrap problem \cite{Friedan:2013cba,Hellerman:2009bu}. Given the torus partition function of a 2d CFT $Z(\beta)\equiv Z(\tau=i\beta,\bar{\tau}=-i\beta)$, modular invariance implies that 
\begin{align}
Z(\beta)&=\chi_0(\beta)+\sum_{\Delta}\rho_\Delta \chi_\Delta(\beta)\\
&=\chi_0(1/\beta)+\sum_{\Delta}\rho_\Delta \chi_\Delta(-1/\beta),
\end{align}
where we have separated the vacuum contribution to the partition function. The sum is over the primary operators, $\beta>0$, $\rho_\Delta$ are positive integers and the characters are given by the product of holomorphic and anti-holomorphic Virasoro characters
\begin{align}
\chi_0(\beta)&=\frac{e^{-2\pi \beta\Delta_0}(1-e^{-2\pi\beta})^2}{|\eta(i\beta)|^2},\\
\chi_\Delta(\beta)&=\frac{e^{-2\pi \beta(\Delta+\Delta_0)}}{|\eta(i\beta)|^2},
\end{align}
where $\Delta_0\equiv -\frac{c-1}{12}$ , $c$ is the central charge of the CFT and $\eta$  denotes the Dedekin $\eta$-function. It is convenient to define the reduced partition function $\hat{Z}(\beta)\equiv |\eta(i\beta)|^2 |i\beta|^{1/2} Z(\beta)$ which satisfies the relation $\hat{Z}(\beta)=\hat{Z}(1/\beta)$. The crossing equation can then be expressed as
\begin{align}\label{splesscrossing}
\mathbb{W}_0+\sum_{\Delta}\rho_\Delta \mathbb{W}_\Delta=0,
\end{align}
with $\mathbb{W}$ defined by
\begin{align}
\mathbb{W}_0&=\beta^{1/2}e^{-2\pi \beta\Delta_0}(1-e^{-2\pi\beta})^2- \beta^{-1/2}e^{-\frac{2\pi}{\beta}\Delta_0}(1-e^{-\frac{2\pi}{\beta}})^2,\\
\mathbb{W}_\Delta&=\beta^{1/2}e^{-2\pi \beta(\Delta+\Delta_0)}-\beta^{-1/2}e^{-\frac{2\pi}{\beta}(\Delta+\Delta_0)}.
\end{align}
The following is a convenient choice of linear functionals
\begin{align}
\mathcal{F}_\Delta^k&=\left.\frac{1}{2(2k-1)!}\left[\frac{(1+\beta)^2}{2}\partial_\beta\right]^{2k-1}\frac{1+\beta}{2\sqrt{\beta}}\mathbb{W}_\Delta\right|_{\beta=1}\\
&=e^{-2\pi (\Delta+\Delta_0)}L_{2k-1}\left(4\pi\left(\Delta+\Delta_0\right)\right),
\end{align}
where $L_{2k-1}$ are odd-indexed Laguerre polynomials. Similarly, for the vacuum contribution we find 
\begin{align}
\mathcal{F}^k_0=e^{-2\pi \Delta_0}\left(L_{2k-1}(4\pi \Delta_0)-2 e^{-2\pi}L_{2k-1}(4\pi (\Delta_0+1))+e^{-4\pi}L_{2k-1}(4\pi (\Delta_0+2))\right).
\end{align}
Acting these functionals on the crossing equations \eqref{splesscrossing} we obtain
\begin{align}
\mathcal{F}^i_0+\sum_\Delta \rho_\Delta \mathcal{F}^i_\Delta=0.
\end{align}
The goal now is to find $\alpha_i$ with $i=1,...,\Lambda$, such that 
\begin{align}
\alpha_i\mathcal{F}^i_0&>0,\\
\alpha_i\mathcal{F}^i_\Delta&\ge0\\
\end{align}
\begin{figure}
	\center\includegraphics[scale=0.8]{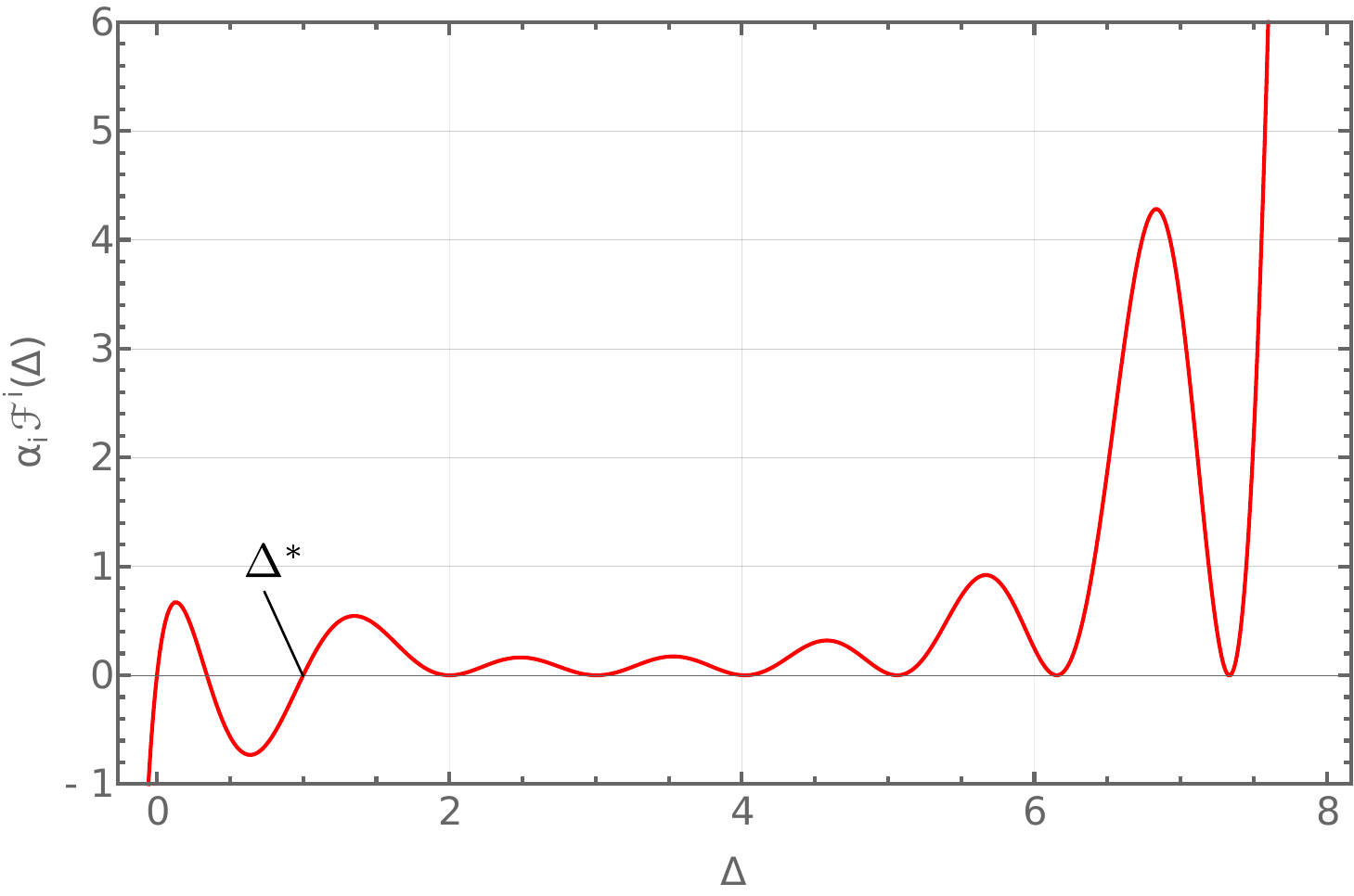}\\
	\caption{Extremal functional for the spinless bootstrap at central charge $c=4$ and number of functionals $\Lambda=10$.}
	\label{fig:functional}
\end{figure}
for all $\Delta\ge\Delta^*$. Furthermore we will minimize over $\Delta^*$ in order to find the extremal functional. This can be carried out employing semidefinite programming as well as the bisection method\footnote{The naviagor function method \cite{Reehorst:2021ykw} can be used to improve convergence in this step, though we have not implemented it in this work.} in order to find  $\Delta^*$ as well as the extremal functional. An example of an extremal functional is plotted in figure \ref{fig:functional}. In \cite{Afkhami-Jeddi:2019zci} it was observed that the extremal functional satisfies the following conditions: 1) The extremal functional is polynomial with a single zero at $\Delta^*$. 2) The extremal functional contains a number of double zeros at values $\Delta>\Delta^*$. 3) The number of double zeros is always equal to $\frac{\Lambda}{2}-1$. Furthermore it was observed that in cases where $\Delta^*$ approached the scalar gap of a known theory, the double zeros rapidly converged to the scaling dimension of other operators in the theory. This motivates an ansatz parameterizing the extremal functional in terms of it's zeros. Using this ansatz we can express this optimization problem in terms of the following Lagrangian
\begin{align}\label{lag1}
\mathcal{L}=\alpha_i\mathcal{F}^i_0+\sum_{\mu=2}^{\Lambda/2} c_\mu\alpha_i\mathcal{F}^i_{\Delta_\mu}+\sum_{\nu=1}^{\Lambda/2} \tilde{c}_\nu\alpha_i\mathcal{F}'^i_{\Delta_\nu},
\end{align}
where $\tilde{c}_\mu$ and $c_\mu$ are Lagrange multipliers imposing the roots of the polynomial and $\mathcal{F}'$ denote the derivative with respect to $\Delta$. Notice that if we dualize the problem, assuming that the extremal functional does not contain any higher order zeros we find the truncated crossing equations,
\begin{align}
\mathcal{F}^i_0+\sum_{\mu=2}^{\Lambda/2} c_\mu\mathcal{F}^i_{\Delta_\mu}=0.
\end{align}
This provides an intuitive explanation of why the roots of the functional provide a good approximation to the low-lying spectrum of the CFT. Furthermore, notice that in this simple case solving the truncated crossing equations automatically yields the optimal solution. This is due to the fact that the number of zeros are such that $\alpha_i$ are uniquely fixed up to a rescaling. More explicitly $\alpha_i$ are given by the kernel of a $2\Lambda\times2\Lambda-1$ matrix
\begin{align}
\vec{\alpha}\propto\ker \left(\begin{array}{c}
\vec{\mathcal{F}}_{\Delta_\mu} \\ 
\vec{\mathcal{F}}'_{\Delta_\nu}
\end{array} \right).
\end{align}
Therefore, it suffices to find the extremal functional, it suffices to solve the truncated crossing equations.
As we will see, in the more general case where additional quantum numbers are present, the number of zeros of the extremal functional do not match the number of equations so nicely and we will be forced to optimize \eqref{lag1} with respect to all variables.
\subsection{Spinning Modular Bootstrap}
The more general problem of spinning modular bootstrap has yielded both numerical and analytic results \cite{Collier:2017shs,Friedan:2013cba,Qualls:2013eha,Hellerman:2010qd, Keller:2012mr,Afkhami-Jeddi:2017idc,Das:2017vej,Apolo:2017xip,Benjamin:2016fhe,Anous:2018hjh,Benjamin:2020zbs,Lin:2021udi}. Let us now briefly review the modular bootstrap set up. We begin by writing the torus partition function\footnote{For simplicity, we will consider CFTs with no currents}
\begin{align}
Z(\tau,\bar{\tau})&=\chi_0(\tau)\bar{\chi}_0(\bar{\tau})+\sum_{h,\bar{h}}\rho_{h,\bar{h}} \chi_{h,\bar{h}}(\tau)\bar{\chi}_{h,\bar{h}}(\bar{\tau})\\&=Z(-1/\tau,-1/\bar{\tau}),
\end{align}
where $h$ and $\bar{h}$ denote the holomorphic and anti-holomorphic weights of primary operators, $\rho_{\Delta,s}\in \mathbb{N}$ are the degeneracies, the second line follow from modular invariance of the torus partition function and $\chi$ and $\bar{\chi}$ are the Virasoro characters
\begin{align}
\chi_0(\tau)&=\frac{e^{2\pi i \tau(-\frac{c-1}{24})}(1-e^{2\pi i \tau})}{\eta(\tau)},\\
\chi_{h}(\tau)&=\frac{e^{2\pi i \tau(h-\frac{c-1}{24})}}{\eta(\tau)}.
\end{align}
It is again convenient to introduce the reduced partition function
\begin{align}\label{modcrossing}
\hat{Z}(\beta,\bar{\beta})&=|\beta|^{1/2}|\eta(i\beta)|^2 Z(i\beta,-i\bar{\beta}).\nonumber\\
&=\hat{Z}(1/\beta,1/\bar{\beta})
\end{align}
Focusing on the parity invariant part of the spectrum we can decompose this in terms of the reduced characters
\begin{align}
\hat{Z}(\beta,\bar{\beta})=\hat{\chi}_0(\beta)\hat{\bar{\chi}}_0(\bar{\beta})+\sum_{\Delta,s}\rho_{\Delta,s}\left(\hat{\chi}_{\frac{\Delta+s}{2}}(\beta)\hat{\bar{\chi}}_{\frac{\Delta-s}{2}}(\bar{\beta})+\hat{\chi}_{\frac{\Delta-s}{2}}(\beta)\hat{\bar{\chi}}_{\frac{\Delta+s}{2}}(\bar{\beta})\right),
\end{align}
where $\Delta=h+\bar{h}$, $s=|h-\bar{h}|$ and the reduced characters are given by
\begin{align}
\hat{\chi}_0(\beta)&=\beta^{1/4}e^{-2\pi \beta(-\frac{c-1}{24})}(1-e^{-2\pi \beta}),\\
\hat{\chi}_{x}(\beta)&=\beta^{1/4}e^{-2\pi  \beta(x-\frac{c-1}{24})}.
\end{align}
Proceeding as before we define the differential operator 
\begin{align}
\mathcal{D}^k&=\left.\frac{1}{2 \Gamma (k+1)}\left[\frac{(1+\beta)^2}{2}\partial_\beta\right]^{2k-1}\left(\frac{\beta +1}{\sqrt{2}}\right)^{1/2} \beta^{-1/4}\right|_{\beta=1}\\
\end{align}
and construct the linear functionals
\begin{align}
\mathcal{F}^{k,\bar{k}}_{\Delta,s}&=\mathcal{D}^k \bar{\mathcal{D}}^{\bar{k}}(
\hat{\chi}_{\frac{\Delta+s}{2}}(\beta)\hat{\bar{\chi}}_{\frac{\Delta-s}{2}}(\bar{\beta})+
\hat{\chi}_{\frac{\Delta-s}{2}}(\beta)\hat{\bar{\chi}}_{\frac{\Delta+s}{2}}(\bar{\beta})
-(\beta \leftrightarrow 1/\beta))\nonumber\\
&=e^{-4\pi (\Delta-\frac{c-1}{12})}L^{-1/2}_{2k-1}\left(4 \pi  \left(\frac{\Delta +s}{2}-\frac{c-1}{24}\right)\right)L^{-1/2}_{2\bar{k}-1}\left(4 \pi  \left(\frac{\Delta -s}{2}-\frac{c-1}{24}\right)\right)\nonumber\\ 
&~+k\leftrightarrow \bar{k},
\end{align}
where $k+\bar{k}$ are odd integers and $L_p^\alpha$ are the generalized Laguerre polynomials. The vacuum contribution $\mathcal{F}^{k,\bar{k}}_{0,0}$ is similarly obtained by acting the differential operators on the reduced vacuum characters. Acting with these functionals on the crossing equation \eqref{modcrossing} we find
\begin{align}
\mathcal{F}^i_{0,0}+\sum_{\Delta,s} \rho_{\Delta,s}\mathcal{F}^i_{\Delta,s}=0,
\end{align}
where we write $i$ as a shorthand for the index of a set $\{(k,\bar{k})|~k+\bar{k}\in 2\mathbb{N}-1~\wedge~\bar{k}<k \}$. We now consider the scalar gap problem by finding a set of numbers $\alpha_i$ such that
\begin{align}\label{modbscond}
\alpha_i\mathcal{F}^i_{0,0}&>0\nonumber\\
\alpha_i\mathcal{F}^i_{\Delta,s}&\ge 0,~~ \begin{cases}
      \Delta>\Delta^* &, s=0\\
      \Delta>s &, s>0
    \end{cases}       .
\end{align}
To find the extremal functional we find the smallest $\Delta^*$ such that the above conditions are satisfied. Unlike the spinless modular bootstrap, we now have many extremal functionals labeled by $s$, each of which may now contain single and double zeros. We can again use this observation to make an ansatz and write this as an optimization problem with the following Lagrangian
\begin{align}
\mathcal{L}=\alpha_i \mathcal{F}_{0,0}^i+\sum_{a\in Q_s} c_a\alpha_i\mathcal{F}^i_a
+\sum_{b\in Q_d} (c_b\alpha_i\mathcal{F}^i_b+\tilde{c}_b\alpha_i\mathcal{F}'^i_b),
\end{align}
where $c_a$,$c_b$ and $\tilde{c}_b$ are Lagrange multipliers imposing the position of the roots of the extremal functional, and $Q_s$ and $Q_d$ are the list of scaling dimensions and spins of operators corresponding to the single and double roots respectively. As previously mentioned the total number of zeros appearing in the extremal functional no longer matches the number of equations in such a way as to trivially fix the coefficients $\alpha_i$\footnote{ We can still directly solve for $\alpha$'s by finding $\ker \left(\begin{array}{c}
\vec{\mathcal{F}}_{\Delta_{\text{single}}} \\ 
\vec{\mathcal{F}}'_{\Delta_{\text{double}}}
\end{array} \right)$,\label{solvealpha} where we keep only enough roots such that the matrix has $\Lambda-1$ rows. However in this work we will treat $\alpha$'s as unknown variables to solve for.}. We must therefore optimize this Lagrangian with respect to all variables simultaneously. Variation with respect to the double roots as well as $\tilde{c}_b$ we find that we must set $\tilde{c}_b$ to zero in order to avoid higher order roots.
Note however that variation with respect to the single roots results in an equation imposing that the extremal functional contains double zeros at these locations. This is inconsistent with our ansatz. We therefor take the location the single roots as fixed parameters of the optimization problem and optimize with respect to all other variables\footnote{The single roots may be treat the  as variables if we fix $N_s$ of the $\alpha$ variables using the equation described in footnote \ref{solvealpha}. However for the results in this paper we treated all single roots except the first scalar as constants and varied other parameters.}. Imposing these conditions the variation yields
\begin{align}\label{modvary}
0&=\mathcal{F}_{0,0}^i+\sum_{a\in Q_s} c_a\mathcal{F}^i_a+ \sum_{b\in Q_d} c_b\mathcal{F}^i_b,\nonumber\\
0&=\alpha_i\mathcal{F}^i_a,~~~~~~~a\in Q_s\nonumber\\
0&=\alpha_i\mathcal{F}^i_b,~~~~~~~b\in Q_d\nonumber\\
0&=\alpha_i\mathcal{F}'^i_b,~~~~~~~b\in Q_d
\end{align}
Choosing a set of functionals labeled by $i=1...\Lambda$ we have $\Lambda+N_s+2N_d$ equations where $N_s$ and $N_d$ count the number of single and double roots. In addition, we are optimizing with respect to the double roots $\Delta$, the coefficients $c$ and $\alpha$ for a total of $\Lambda+N_s+2 N_d$ variables. Since the overall scale of the extremal functional is not importan, we may eliminate one variable by fixing the normalization by setting $\alpha_1=0$. We can then use this extra equation to treat one of the single roots as a variable. In this case, we choose to vary the single root corresponding to the scalar gap. The result is a set of nonlinear equations for the CFT data as well as the coefficients of the extremal functional. Notice that as presented, we have not imposed any positivity/unitarity conditions on the equations \eqref{modvary}. We might therefore hope that the solutions to these equations might have some bearing on non-unitary theories which are typically not accessible through conventional bootstrap methods. 

At this point, we can try to solve these equations numerically using the Newton's method. However, due to the non-linear nature of the equations, convergence is contingent on a good initial guess for the location of the roots and the discrete choice of spins. If we restrict ourselves to unitary theories we can use semidefinite programming to find a good initial guess for the initial conditions of Newton's method. We find that the extremal solutions found using the traditional bootstrap method always yield a solution to \eqref{modvary}. To achieve this, we begin by imposing \eqref{modbscond} and using bisection to find the optimum $\Delta^*$. We then optimize by minimizing that vacuum contribution and extracting the roots of the extremal functionals obtained by this procedure\footnote{Throughout this paper we will utilize the semidefinite program solver SDPB \cite{Simmons-Duffin:2015qma,Landry:2019qug} to find initial conditions for the Newton's method.}. Since $\alpha$ and $c$ appear linearly in \eqref{modvary}, convergence to the optimal solution is not very sensitive to the initial conditions for these variables.

\subsection{Correlation Function Bootstrap}
 We will now apply the techniques described above to bootstrapping CFT 4-point correlation functions on a sphere. We will consider the correlation function of four identical scalar operators for simplicity, but we expect these methods to be generalizable to mixed correlator bootstrap. The optimization problem we find is very similar to that of the modular bootstrap. We begin by writing the crossing equation 
 \begin{align}
\langle \phi(0)\phi(z,\bz)\phi(1)\phi(\infty)\rangle&=\sum_{\Delta,s} \lambda_{\phi\phi\O_{\Delta,s}}^2\frac{g_{\Delta,s}(z,\bz)}{(z\bz)^{\Delta_\phi}}\nonumber\\
&=\sum_{\Delta,s}\lambda_{\phi\phi\O_{\Delta,s}}^2 \frac{g_{\Delta,s}(1-z,1-\bz)}{((1-z)(1-\bz))^{\Delta_\phi}},
 \end{align}
 where $\lambda$ denote the OPE coefficients and $g_{\Delta,s}$ are the conformal blocks for the exchanged operator with the corresponding quantum numbers. Proceeding in the same way as before, acting with linear functionals on the difference of the two channels we obtain
\begin{align}
 \mathcal{F}^i_{0,0}+\sum_{\Delta,s} \lambda^2_{\Delta,s}\mathcal{F}^i_{\Delta,s}=0,
\end{align}
where $\mathcal{F}^i$ are defined by evaluating coordinate derivatives at the symmetric point 
\begin{align}
\mathcal{F}^i_{\Delta,s}=\partial^k\bar{\partial}^{\bar{k}} \left[((1-z)(1-\bz))^{\Delta_\phi}g_{\Delta,s}(z,\bz)-(z\bz)^{\Delta_\phi}g_{\Delta,s}(1-z,1-\bz)\right]_{z=\bz=1/2},
\end{align}
and $i$ labels the pair of integers $k$ and $\bar{k}$ drawn from a set such that their sum is odd. The conformal blocks can be computed in closed form in even dimensions \cite{Dolan:2000ut,Dolan:2003hv,Dolan:2011dv}. However, we will make use of a recursion relation\cite{Kos:2013tga,Penedones:2015aga,Zamolodchikov:1984eqp} \footnote{We use a C++ implementation of the recursion relations which can be found at \href{https://gitlab.com/bootstrapcollaboration/scalar_blocks}{$https://gitlab.com/bootstrapcollaboration/scalar\_blocks$ }} for computing the derivatives of conformal blocks in general spacetime dimensions. To probe the parameter space of unitary CFTs we again try to construct extremal functionals by imposing
\begin{align}
\alpha_i\mathcal{F}^i_{0,0}&>0,\nonumber\\
\alpha_i\mathcal{F}^i_{\Delta,s}&\ge 0,~~ \begin{cases}
      \Delta>\Delta^* &, s=0\\
      \Delta>s+d-2 &, s>0
    \end{cases}    
\end{align}
where the inequality for the non-scalar operators follows from unitarity. We then find the smallest $\Delta^*$ for which these conditions are satisfied. Recasting this problem in terms of a Lagrangian we find
\begin{align}
\mathcal{L}=\alpha_i \mathcal{F}_{0,0}^i+\sum_{a\in Q_s} c_a\alpha_i\mathcal{F}^i_a
+\sum_{b\in Q_d} (c_b\alpha_i\mathcal{F}^i_b+\tilde{c}_b\alpha_i\mathcal{F}'^i_b),
\end{align}
where $Q$ and $c$ are defined in the same way as described in the previous section. Variation with respect to the variables results in identical equations of motion \eqref{modvary} as described in the preceding section. Using semidefinite programming to find the extremal functionals to use as initial conditions in solving these equations, we can confirm that the procedure always converges to a solution. Note that the first equation in \eqref{modvary} is the truncated crossing equation. We can therefore interpret the Lagrange multipliers as an approximation to the OPE coefficients in the same way that the roots approximate the scaling dimensions.
\section{Deforming the Solutions} \label{dts}
\subsection{Modular Bootstrap}
\subsubsection{Unitary Deformation}

So far we have described a procedure for extracting CFT data and verifying the extremal solutions obtained using semi-definite programming. We now show that by deforming the solutions we can access other regions in the parameter space of CFTs. To be concrete, let us start by fixing the central charge of the CFT at $c=3.5$ and find the extremal functional at $\Lambda=6$. We will then use this solution as initial conditions to \eqref{modvary} while varying other parameters. We will choose to vary the central charge in this case. This procedure allows us to keep track of the scaling dimensions and the degeneracies along the deformation. 
\begin{figure}
	\center\includegraphics[scale=0.8]{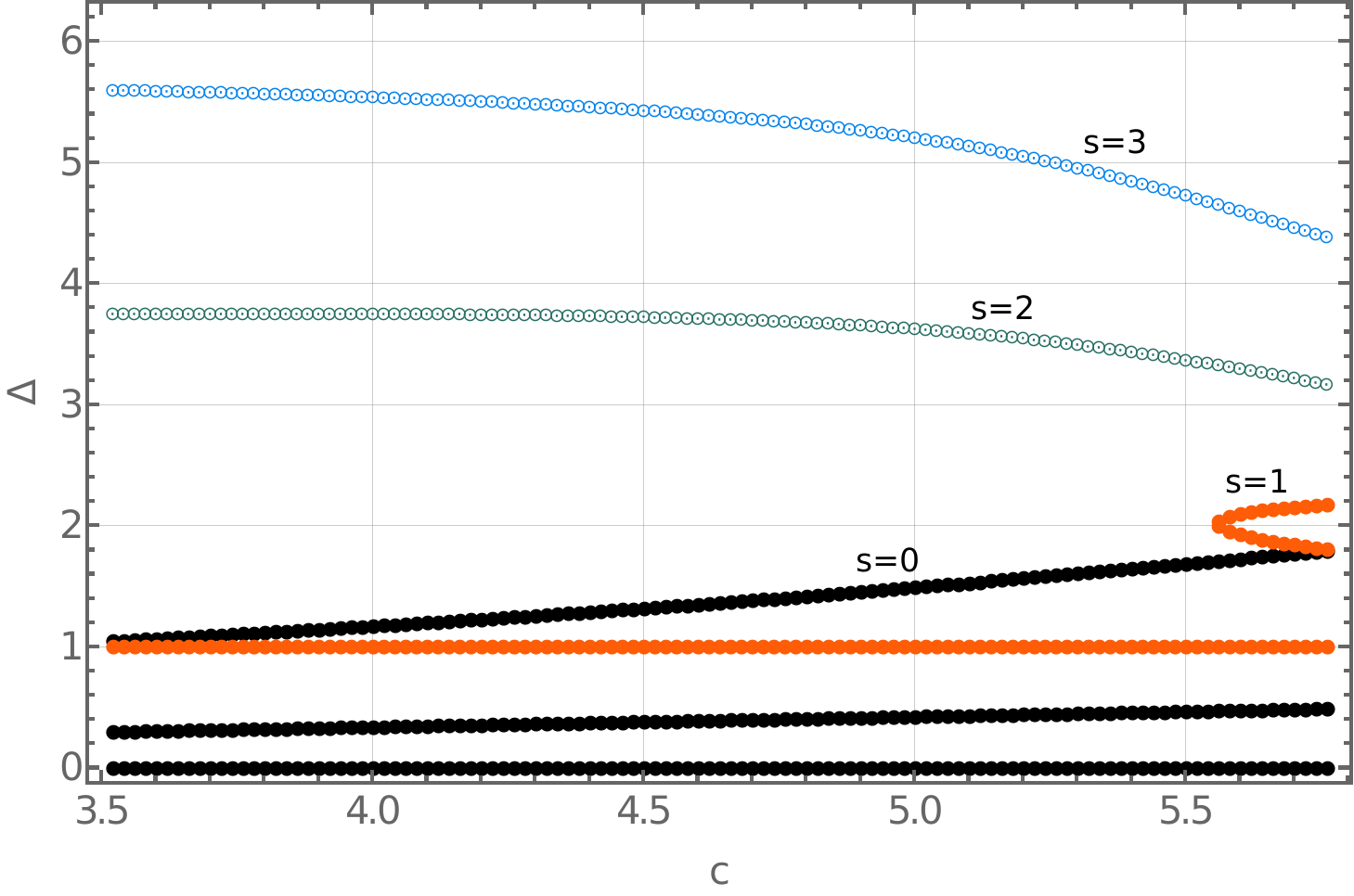}\\
	\caption{Roots of the extremal functional evaluated at increasing values of the central charge. Single roots are denoted by filled circles and double roots are denoted by dotted circles.}
	\label{fig:modflow1}
\end{figure}

\begin{figure}
	\center	\includegraphics[scale=0.8]{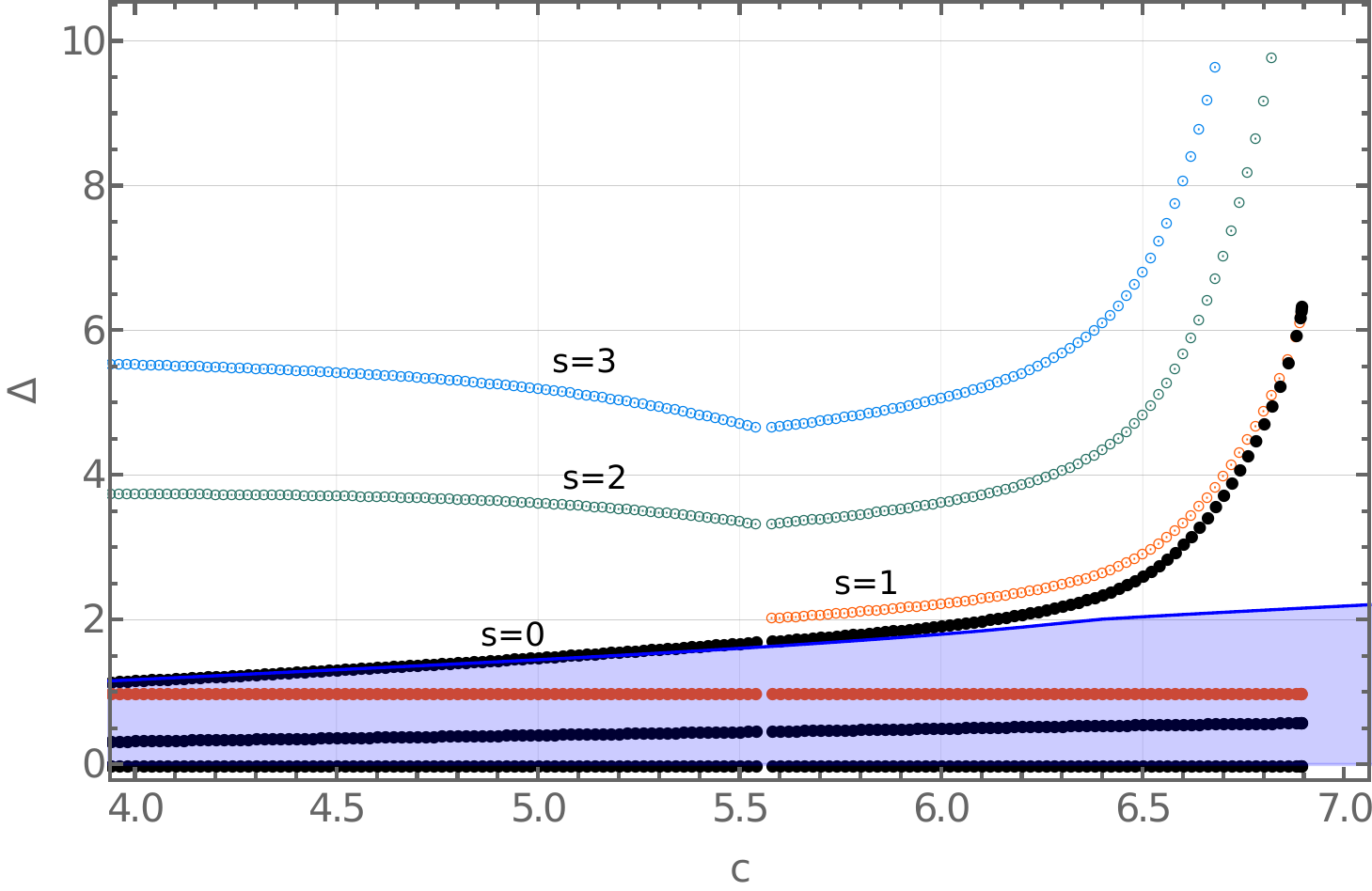}\\
	\caption{Roots of the extremal functional evaluated at increasing values of the central charge after the addition of a double root at spin 1. Single roots are denoted by filled circles and double roots are denoted by dotted circles. The blue line shows the extremal gap obtained by semi-definite programming.}
	\label{fig:modflow2}
\end{figure}

It is also useful to explicitly construct the functionals as we deform the solution. We plot the roots of the functional along the deformation in figure \ref{fig:modflow1}. By studying these roots we see that new single roots appear in the functional at spin 1 as the central charge becomes larger than 5.56. For central charges below this value, the extremal functionals we have constructed place rigorous bounds on the space of CFTs. The appearance of new single roots above the unitarity bound, spoils the straightforward interpretation of the functionals. We can interpret the functional with the new single roots as ruling out CFTs satisfying \eqref{modbscond} by also excluding operators with scaling dimensions between the new single roots. This is due to the fact that the only region where the functional has become negative above the unitarity bound, is between the new roots. Alternatively, we can try to introduce a new double root to the spectrum at the central charge where the single roots first appear and check whether a new solution exists despite the addition of the new operator to the spectrum. We find that adding a new double root to the spectrum results in a new solution. Indeed, we have observed that whenever a pair of new single roots appear in the functionals, the addition of an operator to the spectrum yields a new solution to \eqref{modvary} and so this phenomenon seems to be true more generally. We can then continue to deform our solution after the addition of the new operator while obtaining rigorous bounds on the parameter space of allowed unitary CFTs. The roots of the extremal functional after the addition of a new double root are plotted in figure \ref{fig:modflow2}. We see that the spectrum along the deformation begins to diverge as we approach the kink in the extremal functional obtained using the traditional unitary bootstrap.

\subsubsection{Non-Unitary Deformation}
\begin{figure}
\center	\includegraphics[scale=0.8]{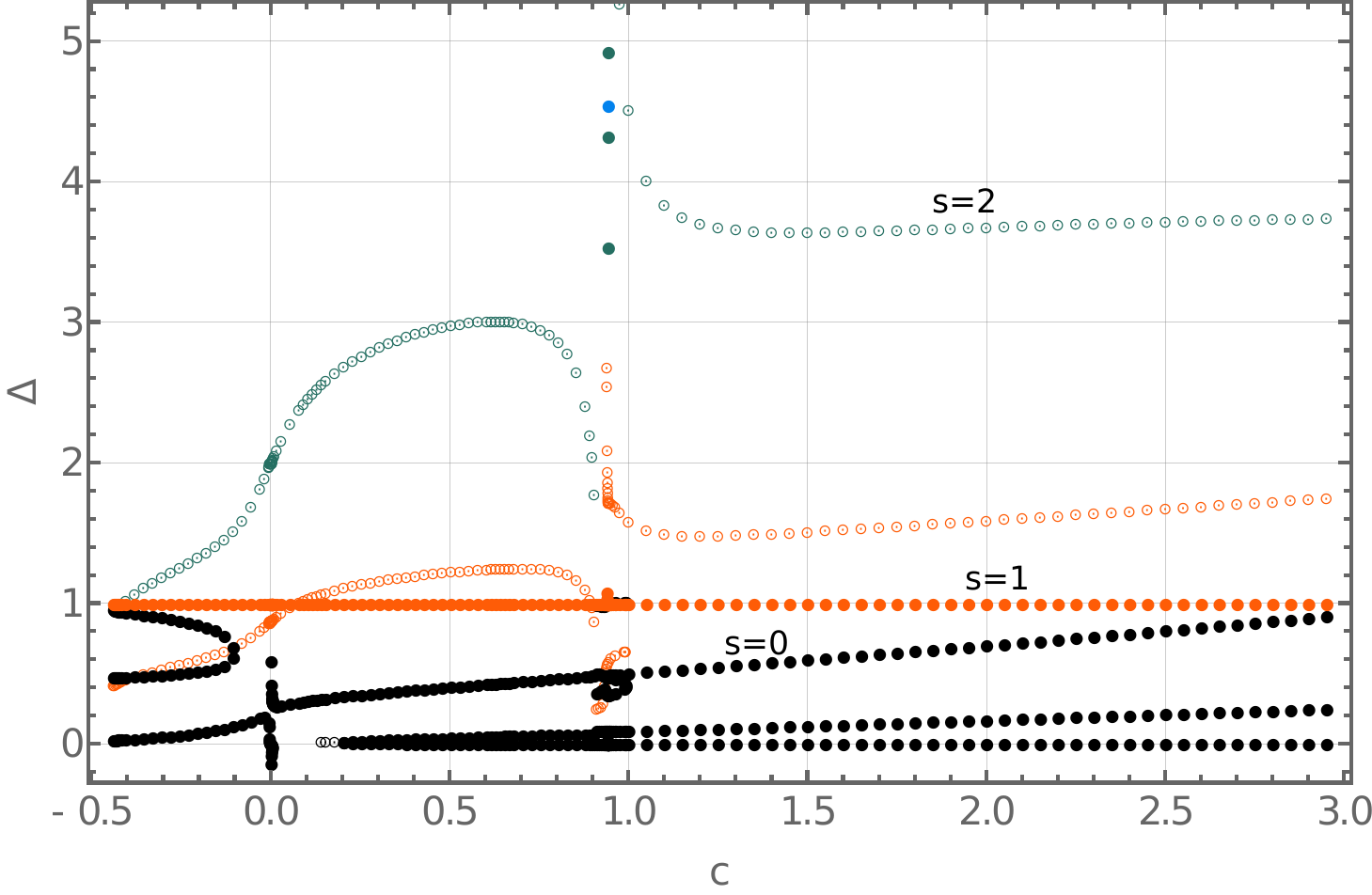}
	\caption{Roots of the extremal functional evaluated at decreasing values of the central charge. Single roots are denoted by filled circles and double roots are denoted by dotted circles. }
	\label{fig:modflow3}
\end{figure}
\begin{figure}
\center	\includegraphics[scale=0.8]{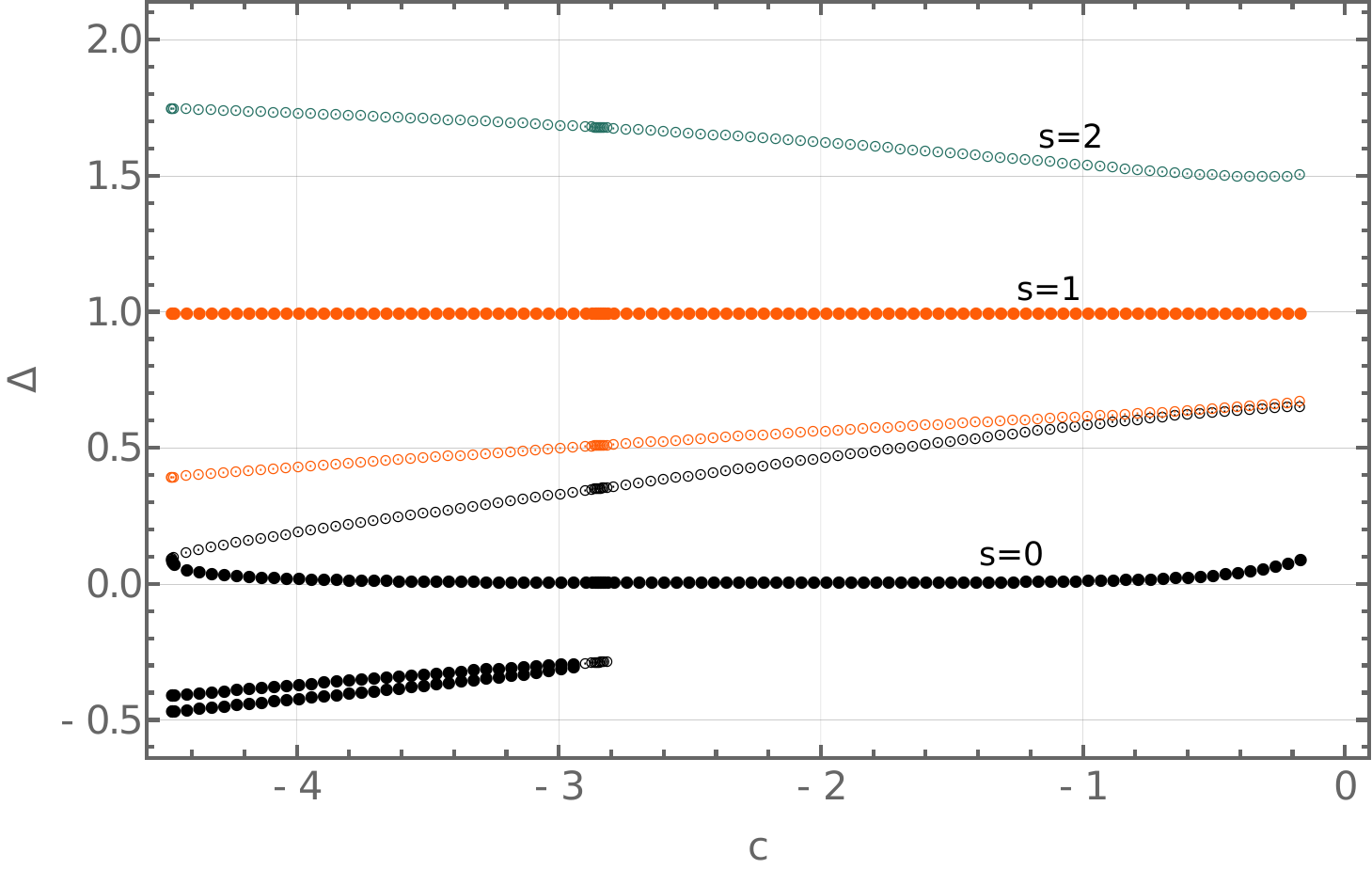}
	\caption{Roots of the extremal functional evaluated at decreasing values of the central charge after addition of a double root at spin 0. Single roots are denoted by filled circles and double roots are denoted by dotted circles.}
	\label{fig:modflow4}
\end{figure}

In the previous section, we described deformations of the extremal solutions for increasing values of the central charge. In this section, we will describe a deformation where we decrease the central charge into negative values. The procedure is the same as above. We start at $c=3$ and decrease the central charge. The roots of the extremal functional along this deformation are plotted in figure \ref{fig:modflow3}. We again encounter the appearance of a pair of roots as we approach $c=-0.12$. We find that the deformation terminates if we continue the deformation without adding a double root to the solution. So we continue by adding a new double root at spin-zero and continue the deformation. The result of this deformation is shown in figure \ref{fig:modflow4}. We find a new pair of roots appearing at $c=-2.8$ that persist as we continue to $c=-22/5$. The position of these roots is near $-2/5$. This value is suggestive since the non-unitary Yang-Lee minimal model contains a scalar operator with matching scaling dimension and central charge. In addition, there are roots at spins 1 and 2 with scaling dimensions numerically close to the theoretical values of $3/5$ and $8/5$ respectively.  As we will see in the next section, operators corresponding to this minimal model are also present in the extremal solutions for correlation funcions in 2d in the non-unitary region of the parameter space. The observation that the Yang-Lee and the Ising extremal solutions may be connected by this type of deformation was noted in \cite{El-Showk:2016mxr}.

\subsection{Correlation Function Bootstrap}
\subsubsection{Deformation in 2d : Ising to Yang-Lee}

In this section, we will use the methods outlined above to deform the extremal solution arising in the study of 4 point function of identical scalars. We begin by considering the 4-point function of operators with scaling dimension $\Delta_\phi=0.155$ in 2 spacetime dimensions. We find the extremal solution utilizing semidefinite programming and use this as initial conditions to solve \eqref{modvary}. We then deform this solution by varying the scaling dimension of the external operator $\Delta_\phi$. We set $\Lambda=8$ for the calculations performed in this section.

\begin{figure}
\center	\includegraphics[scale=0.8]{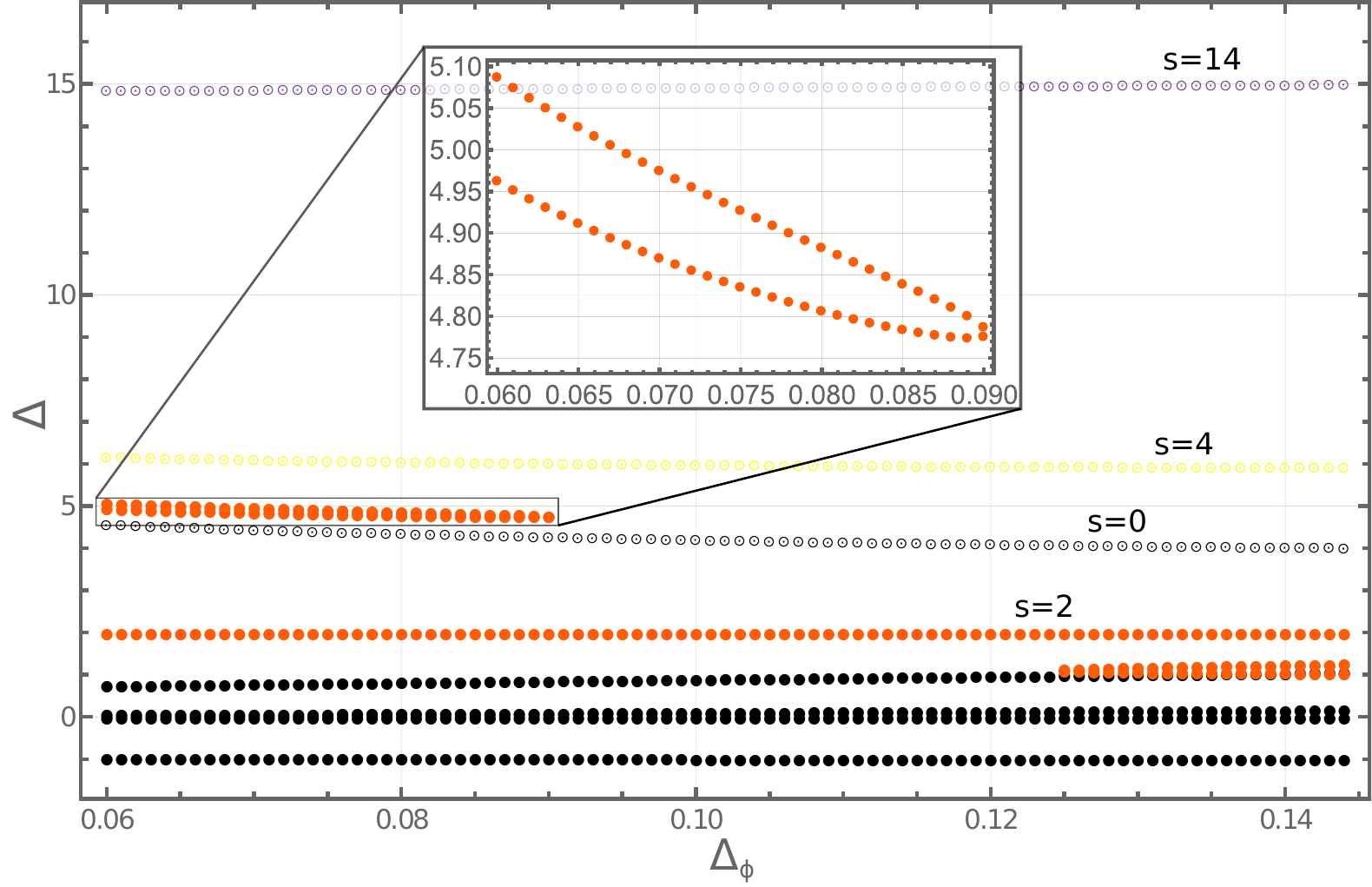}
	\caption{Roots of the extremal functional evaluated for decreasing  $\Delta_\phi$. Single roots are denoted by filled circles and double roots are denoted by dotted circles. The inlay plot shows the appearance of the pair single zeros at spin 2.}
	\label{fig:corrbs1}
\end{figure}

It is well established \cite{Rychkov:2009ij} that the extremal solutions obtained using semidefinite programming result in a curve with a kink that corresponds to the 2d Ising minimal model. The presence of the kink is manifested in the deformed solution by the fact that the squared OPE coefficients of certain operators become negative near the expected location of the kink.. At these points, we can choose to decouple these operators from the spectrum and continue along the unitary branch in which case we will reproduce the well-known bootstrap curve. Alternatively, we can continue the solution as is, and explore the solution that do not necessarily correspond to unitary theories. The roots of the extremal functional along this deformation are shown in figure \ref{fig:corrbs1}. Along this deformation, we see the appearance of a pair of single roots once more. In this case, we see that if we do not add a double root in place of the new roots and keep deforming, the solution eventually terminates and we are unable to proceed to negative values of $\Delta_\phi$. However, by adding a new double root to the spectrum, we can deform the solutions into the non-unitary region of the parameter space.

\begin{figure}
\center	\includegraphics[scale=0.8]{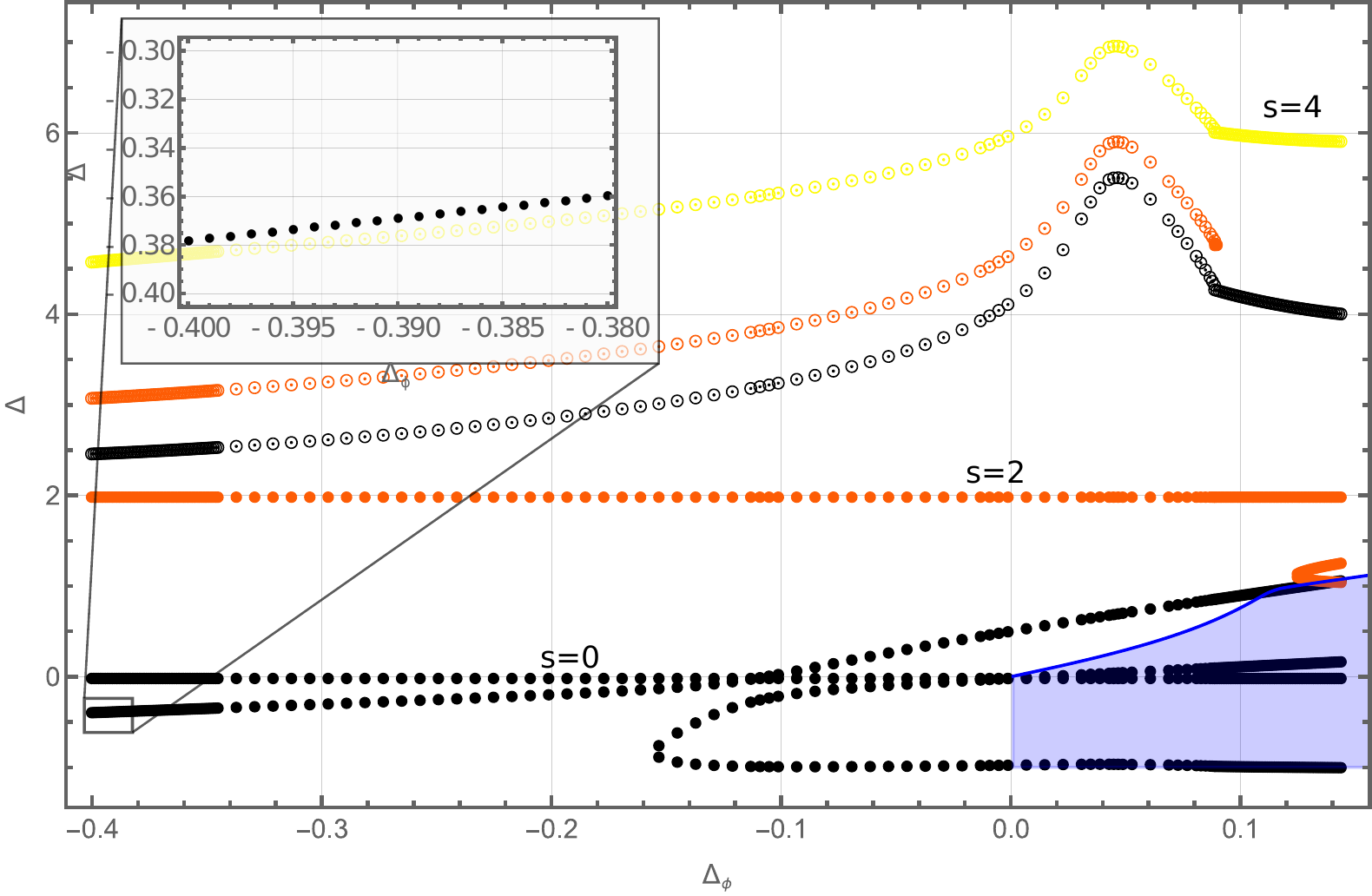}
	\caption{Roots of the extremal functional evaluated at decreasing values of the $\Delta_\phi$ after the addition of a double root at spin 2. Single roots are denoted by filled circles and double roots are denoted by dotted circles. The blue curve Shows the extermal scalar gap obtained using semidefinite programming}
	\label{fig:corrbs2}
\end{figure}

The roots of the extremal functional along this deformation are shown in figure \ref{fig:corrbs2}. Based on this figure we observe that the scalar extremal functional has a root that passes numerically near the point $-2/5$ when the external scaling dimension $\Delta_\phi$ is also $-2/5$. This provides evidence that the extremal solution corresponds to the Yang-Lee minimal model. The spectrum along the deformation contains a spin 2 operator corresponding to the stress-tensor. Using the Ward identity we can extract the central charge of the CFT along the deformation. More specifically we find that at $\Delta_\phi=-2/5$ the central charge $c=-4.7$ which is numerically close to the theoretical value of $c=-22/5$ for the Yang-Lee minimal model providing additional evidence for the claim. In addition, there are two roots in the spin 4 and 14 extremal functionals with scaling dimensions $4.6$ and $14.4$ respectively. These should be compared with the operators of the same spin in the Yang-Lee model which have scaling dimensions $18/5$ and $68/5$ respectively. We expect these values to better converge to the exact values with increasing $\Lambda$.
\subsubsection{Deformation in 3d}
\begin{figure}
\center	\includegraphics[scale=0.8]{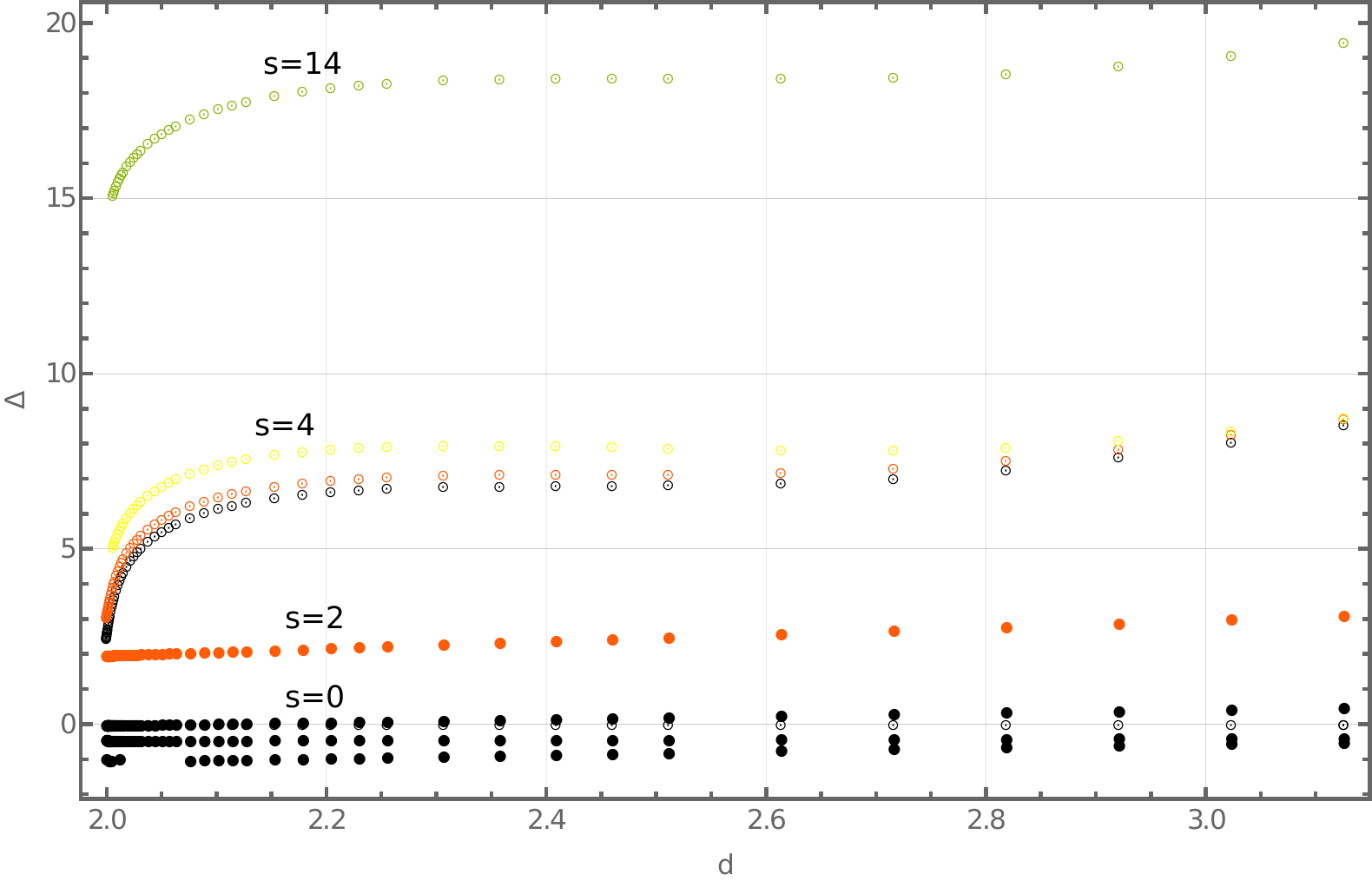}
	\caption{Roots of the extremal functional for increasing spacetime dimension d. Single roots are denoted by filled circles and double roots are denoted by dotted circles. }
	\label{fig:corrbsd}
\end{figure}
Given the results in the previous section, it is natural then to ask whether a similar deformation can give insights into the CFT data of non-unitary theories in higher spacetime dimensions. In the rest of this section, we will carry out such a deformation and compare the CFT data we obtain with those of the $\epsilon$-expansion \cite{Butera:2012tq}. We perform the deformation from two different starting points that we describe below. 

\begin{figure}
\center	\includegraphics[scale=0.8]{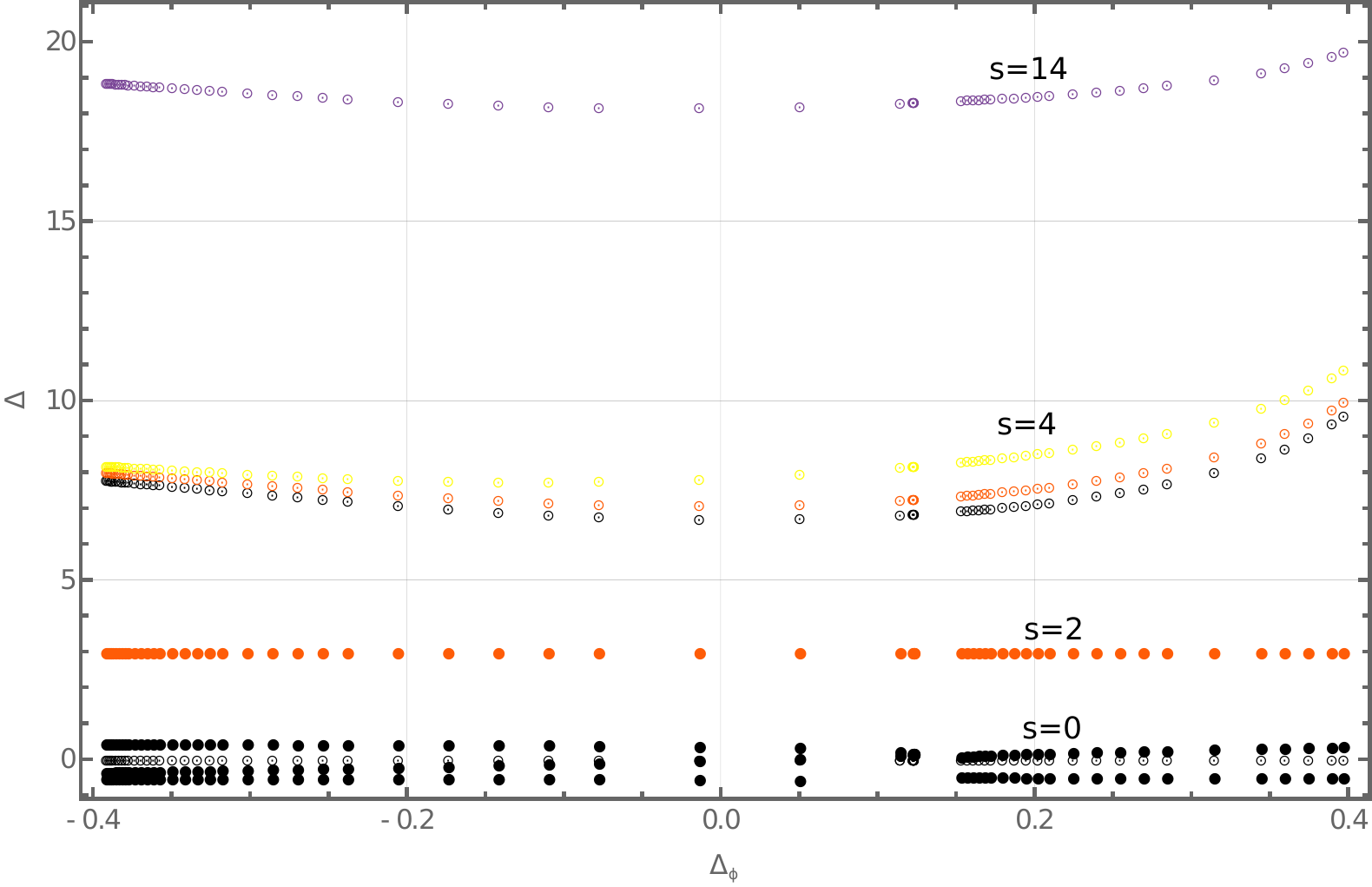}
	\caption{Roots of the extremal functional in 3 dimensions evaluated at increasing values of $\Delta_\phi$. Single roots are denoted by filled circles and double roots are denoted by dotted circles.}
	\label{fig:corrbs3d}
\end{figure}

\begin{figure}
\center	\includegraphics[scale=0.8]{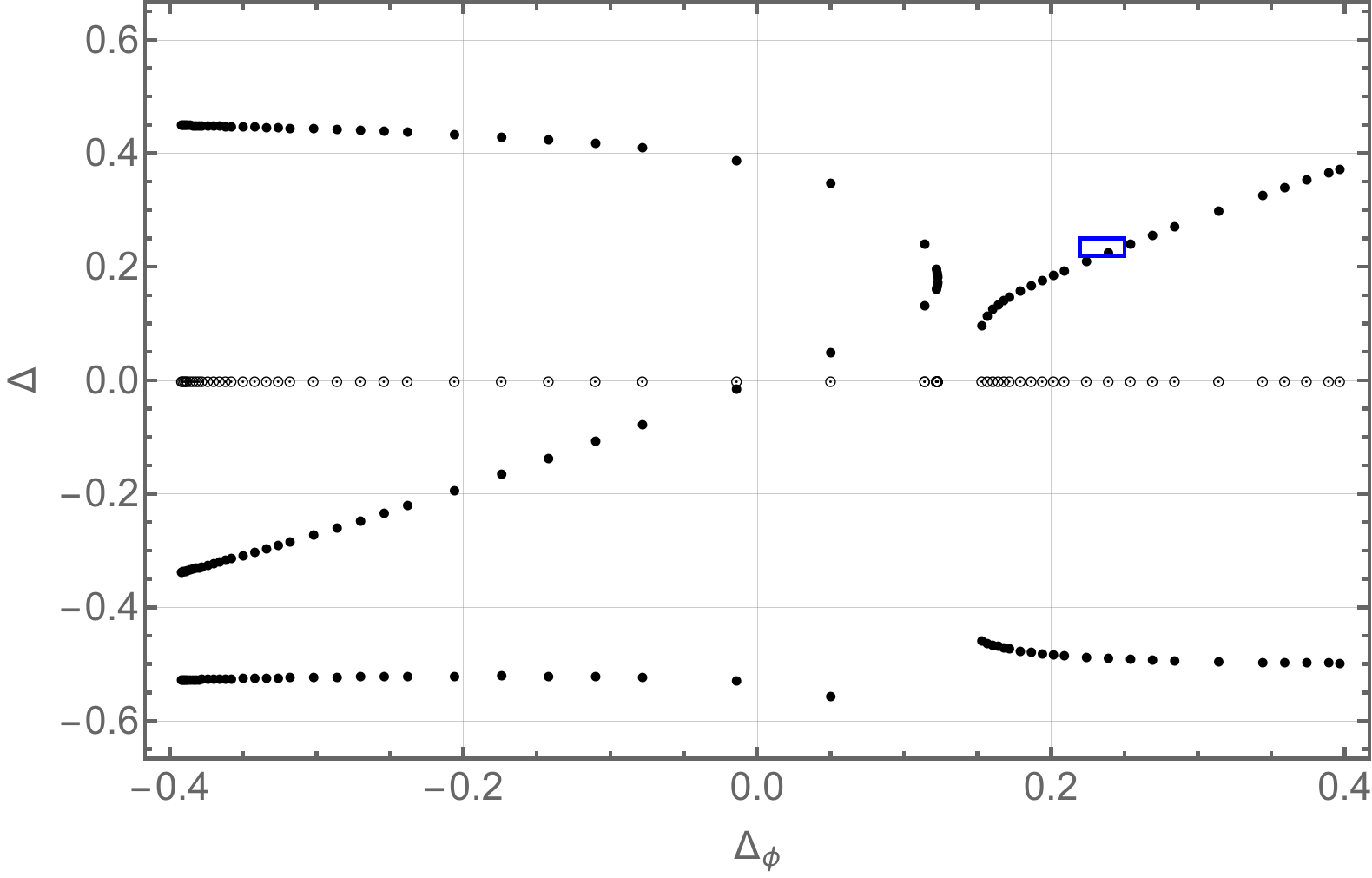}
	\caption{Spin 0 roots of the extremal functional in 3 dimensions evaluated at increasing values of $\Delta_\phi$. The small blue box region is the expected scaling dimension obtained using the $\epsilon$-expansion.  Single roots are denoted by filled circles and double roots are denoted by dotted circles.}
	\label{fig:corrbs3ds}
\end{figure}

First, we will deform the 2d extremal solution corresponding to the Yang-Lee model that was obtained in the previous section from 2 to 3 dimensions. The roots of the functionals along this deformation are shown in figure \ref{fig:corrbsd}. We do not observe the appearance of new roots along this deformation. We then deform this solution by varying the scaling dimension of the external operator $\Delta_\phi$. The roots of the extremal functional along this deformation are shown in figures \ref{fig:corrbs3d} and \ref{fig:corrbs3ds}. We had the advantage of having exact information of a non-unitary minimal model and its scaling dimensions to check against the spectrum along the deformation in two dimensions. This is no longer the case in 3 dimensions. However as can be seen from figure \ref{fig:corrbs3ds}, there indeed exists a root in the scalar extremal functional passing through the window of scaling dimensions predicted using other methods\cite{Butera:2012tq}.

\begin{figure}
\center	\includegraphics[scale=0.8]{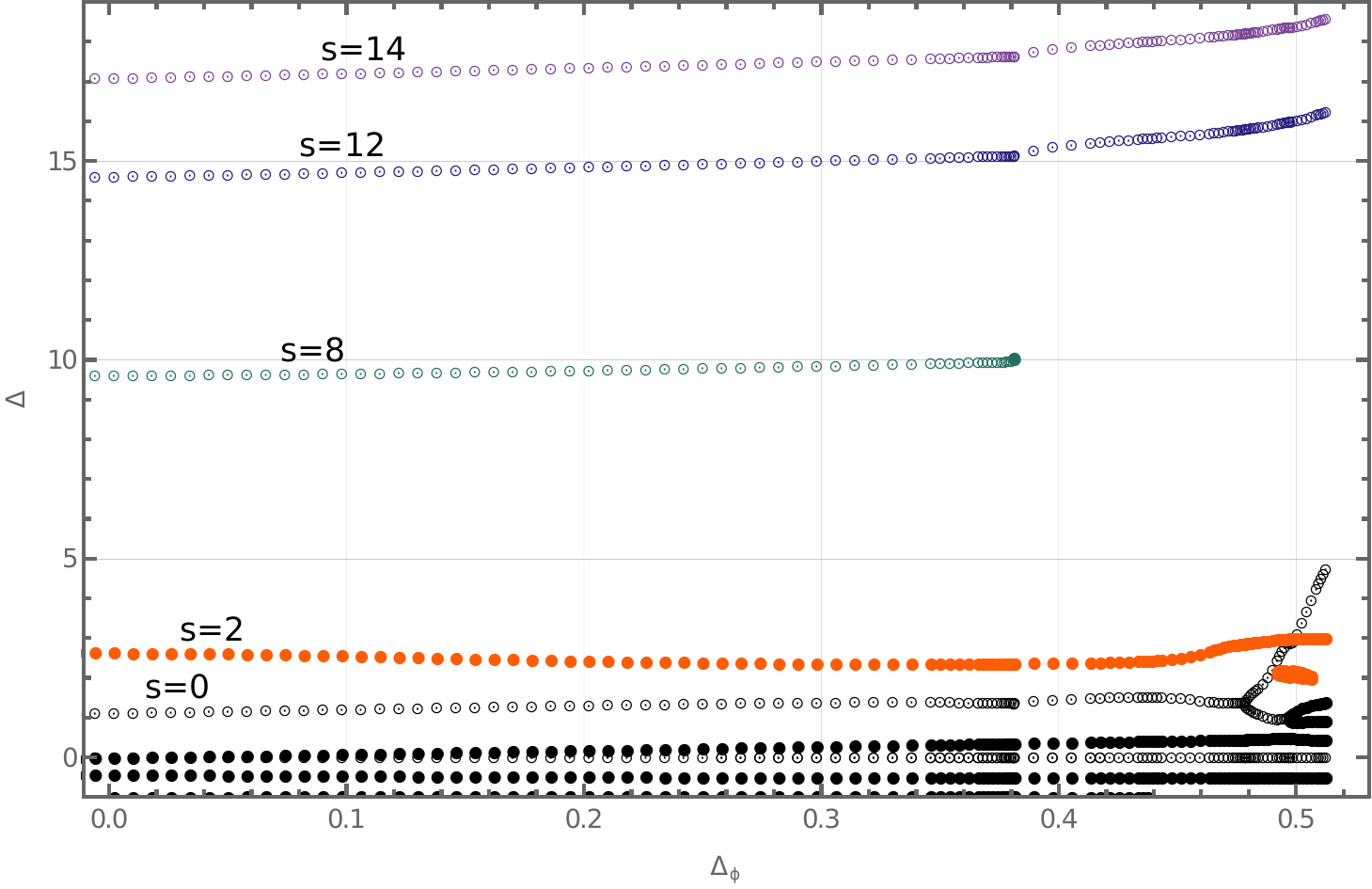}
	\caption{Roots of the extremal functional in 3 dimensions evaluated at decreasing values of $\Delta_\phi$. Single roots are denoted by filled circles and double roots are denoted by dotted circles.}
	\label{fig:corrbs3d2}
\end{figure}

\begin{figure}
\center	\includegraphics[scale=0.8]{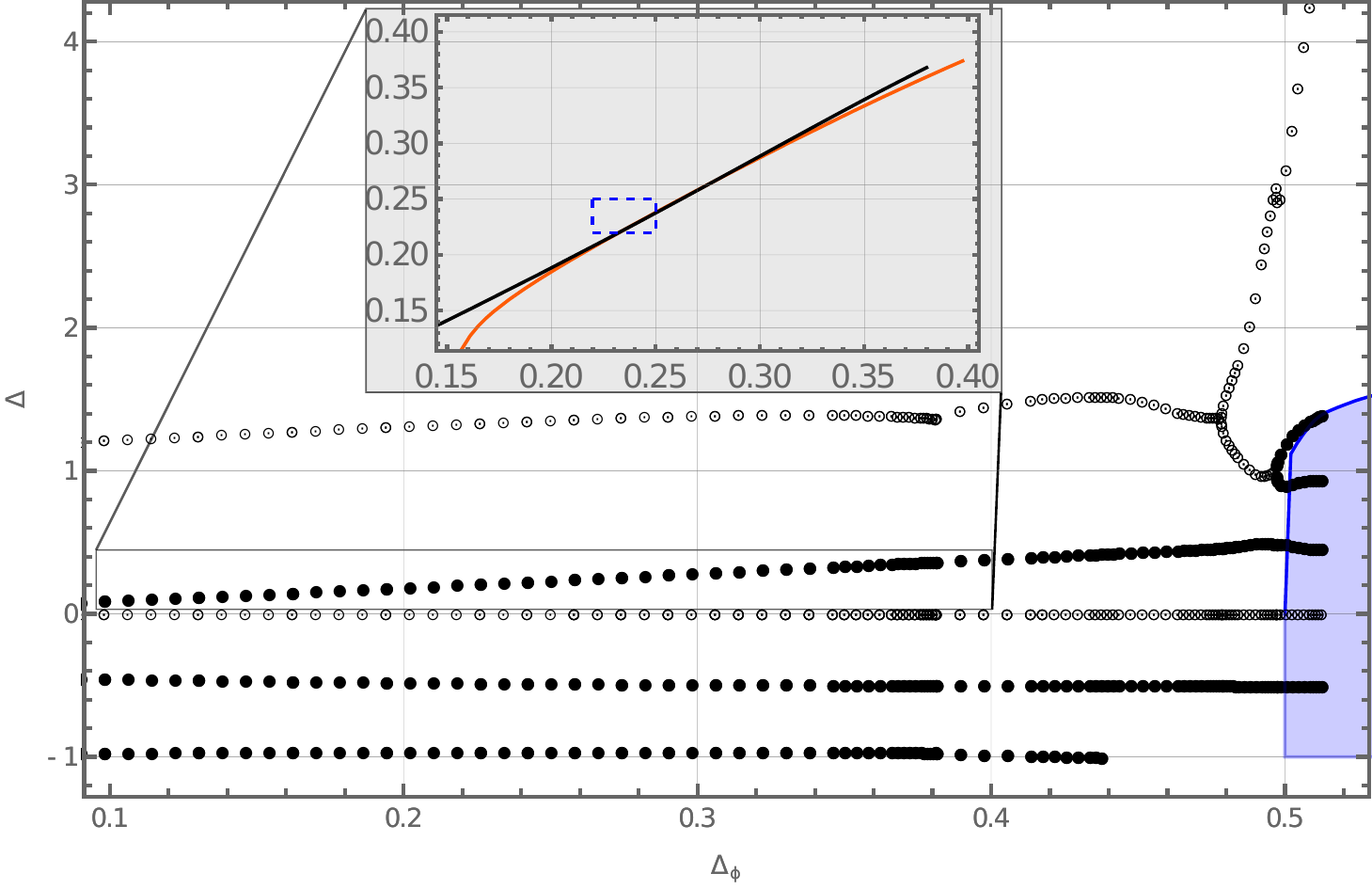}
	\caption{Spin 0 roots of the extremal functional in 3 dimensions evaluated at decreasing values of $\Delta_\phi$. Single roots are denoted by filled circles and double roots are denoted by dotted circles. The Blue line shows the extremal gap obtained by semi-definite programming. The orange line in the subplot shows the spin-zero single root in the deformation discussed in the previous section. The dashed blue region in the inlay is the expected scaling dimension from the $\epsilon$-expansion}
	\label{fig:corrbs3d2sc}
\end{figure}

As the second starting point, we will use semidefinite programming directly applied to the 3-dimensional unitary bootstrap. We start by finding the extremal functional corresponding to $\Delta_\phi=0.52$. As in the 2d case, we deform this theory by decreasing $\Delta_\phi$ and continue past the kink where the squared OPE coefficients switch signs. The roots along this deformation are shown in figure \ref{fig:corrbs3d2} and \ref{fig:corrbs3d2sc}. We do observe the appearance of new pairs of roots along this deformation and as before, we add new operators to the spectrum and continue the deformation. In addition, as shown in figure \ref{fig:corrbs3d2sc}, this deformation exhibits a new feature which is the merging of roots to produce a new double root. Along this deformation, we once again find a scalar operator in the extremal functional intersecting the curve we found previously in the window of scaling dimensions predicted by the $\epsilon$-expansion. This provides non-trivial evidence supporting the belief that the extremal solutions we describe do correspond to a non-unitary theory in 3-dimensions.  This allows us to give an approximate value for the low-lying CFT data within this window. Note that the curves corresponding to the roots do not exhibit features such as kinks. Therefore we do not have a procedure to detect the presence of a non-unitary theory along the deformation. However, crossing of multiple roots in deformations with different starting points may serve as a guide for this purpose.

\section{Discussions and Outlook}\label{dao}
We have demonstrated that the conformal bootstrap may be recast as a constrained optimization problem. Furthermore, we have shown how the solutions of this optimization problem may be deformed, allowing us to explore a broader parameter space than is possible using traditional bootstrap techniques. More specifically, we demonstrated that non-unitary regions in parameter space can be explored using these deformations. In this way, we have provided evidence that different theories corresponding to the solutions of the crossing equation are continuously connected by deformation. It is perhaps surprising that the Ising model and the Yang-Lee model may be connected in this way since the operator contents of these theories are in representation of different symmetry groups. However, it appears that the crossing equations at finite $\Lambda$ are insensitive to this fact.

Finding extremal solutions in this way has multiple advantages. One obvious advantage is the significant computational speed-up in finding extremal boundaries. The computation of each point along the deformation is typically completed in a fraction of a second on a laptop computer. Although we have kept the dimensionality of equations low by restricting to small values of $\Lambda$, we have confirmed that convergence to solutions of \eqref{modvary} occurs within $\sim 20$ seconds for $\Lambda$ as high as 40.

Another advantage of our method derives from the fact that the solutions are parametrized directly in terms of the CFT data. Although we have not carried this out in this paper, this parameterization allows us to choose any of the CFT data as a deformation parameter. It would be interesting to explore this in future work. In addition, this fact allows us to make use of any prior knowledge we may have about the CFT data by incorporating them as fixed parameters into the equations.

Another potential use of this method is its application to systems of mixed correlators which result in island regions in parameter space \cite{Kos:2014bka,Kos:2015mba,Kos:2016ysd}. Due to the presence of multiple channels, these problems typically involve large numbers of equations, resulting in a computationally expensive procedure to map out the island. It is possible to adapt this technique by finding one or two points on the boundary of the island using traditional methods and mapping the rest of the island by deforming the extremal solution along the boundary. This would result in a procedure that is much less computationally demanding. We noted above that we may connect solutions through deformations in spacetime dimensions. It would be interesting to investigate whether the island regions can be deformed across dimensions, as this would allow for the identification of specific regions of interest in the space of extremal solutions.

We chose arbitrary starting points for the initial conditions of the deformations carried out in this paper. We have observed different behavior along the deformation that depends on this choice. Some starting points require the addition of multiple new roots, while others admit no new additions and terminate prematurely or have diverging spectra. In addition, the only reliable source for initial conditions is finding an extremal functional using traditional bootstrap methods, which are limited to unitary regions of the parameter space. It would be interesting if reinforcement-learning techniques utilized in \cite{Kantor:2021jpz,Kantor:2021kbx} can be leveraged to produce initial conditions for deformations. Furthermore, in this work, detection and addition of new roots along the deformation are carried out by hand. This becomes increasingly difficult for larger values of $\Lambda$ as the appearance of roots occurs more frequently. Automating this task would be crucial for exploring the parameter space of extremal functions at larger values of $\Lambda$.

In this paper we have demonstrated that extremal solutions to the bootstrap may be deformed along different directions, including the spacetime dimension, the central charge, or the dimension of the external operators. Deformation with respect to other parameters remains as yet unexplored. It is also not clear whether large, non-trivial deformations are possible for larger values of $\Lambda$ which are typically used in the modern bootstrap. It would be interesting to map out the space of these deformations with an eye towards understanding the conditions under which large deformations are possible. We leave the exploration of these issues to future work.
\section*{Acknowledgements}

We thank Subham Dutta Chowdhury for comments on the manuscript and Shai Chester for sharing the Mathematica code for implementing SDPB. NAJ is supported by the Leo Kadanoff Fellowship and by the US Department of Energy DE-SC0021432. This work was completed utilizing resources provided by the University of Chicago Research Computing Center. This work was performed in part at the Aspen Center for Physics, which is supported by National Science Foundation grant PHY-1607611.

\end{spacing}
\small
\bibliography{deformbib}
\bibliographystyle{ourbst}

\end{document}